\documentstyle[12pt,dina4p]{article}
\begin{document}


\centerline{\normalsize DESY 95--246\hfill ISSN 0418-9833}
\centerline{\normalsize December 1995\hfill hep-ph/9512341}

\vspace{1.0cm}
\begin{center} \begin{Large} \begin{bf}
Extraction of $\alpha$ From the CP Asymmetry
in $\bf {B^{0}/\bar B^0 \rightarrow \pi^{+}\pi^{-}}$ Decays
\end{bf} \end{Large} \end{center}
\vspace{1cm}
\begin{center}
    G.\ Kramer$^a$, W.\ F.\ Palmer$^b$ and Y.\ L.\  Wu$^b$  \\
      \vspace{0.3cm}
        $^a$II. Institut f\"ur Theoretische Physik\footnote{Supported by
            Bundesministerium f\"ur Forschung und Technologie,
            05\,6\,HH\,93P(5), Bonn, Germany and EEC Program
            ``Human Capital and Mobility" Network ``Physics at High Energy
            Colliders" CHRX-CT93-0357 (DG 12 COMA)}\\
            der Universit\"at Hamburg,\\
        D--22761 Hamburg, Germany\\

  \vspace{0.3cm}
        $^b$Department of Physics, The Ohio State University
\footnote{Supported in part by the US
            Department of Energy under contract DOE/ER/01545-605.}, \\
        Columbus, Ohio 43210, USA\\
        \end{center}
  \vspace{1cm}
\noindent {\bf Abstract}\\
\\
The influence of strong and electroweak penguin amplitudes in $B/ \bar B
\rightarrow \pi^+\pi^-$ is investigated in connection with the determination of
the unitarity triangle angle $\alpha$ of the CKM matrix.  A relation
between the observable asymmetry, the angle $\alpha$, and
the penguin amplitude is established.  A model calculation of the penguin
amplitude shows that the CP asymmetry in
$B^{0}\rightarrow \pi^{+}\pi^{-}$ decays is only mildly influenced by the
penguin amplitudes.  Experimental limits on pure penguin and penguin dominated
processes are consistent with the model.  This information also suggests
in a rather model independent way that penguin
amplitudes will not be a serious complicating factor in the determination of
$\alpha$ from the $\pi^{+}\pi^{-}$ time dependent asymmetry.

\parbox[t]{\textwidth}{ }
\newpage

\section{Introduction}

It is expected that B decays will show large CP- violating effects,
characterized by non-vanishing values of the angles $\alpha,~\beta$
and $\gamma$ in the unitarity triangle \cite{NQ}. One of the best ways
to detect this CP violation is to measure an asymmetry between $B^0$
and $\bar {B}^0$ decays into a CP eigenstate. If only one weak
amplitude contributes to the decay, the phase in the elements of
the Cabibbo-Kobayashi-Maskawa (CKM) matrix can be extracted without
uncertainties due to unknown hadronic matrix elements. Thus
$\sin 2\alpha,~\sin 2\beta$ and $\sin 2\gamma$ can in principle be
measured in $B^0,~\bar {B}^0\rightarrow \pi^+\pi^-,~J/\psi K_S$
and $B_s,~\bar {B}_s \rightarrow \rho^0 K_S$ decays, respectively.
Unfortunately the situation is more complicated. In all of the above
cases, in addition to the tree contribution there are amplitudes
due to strong and electroweak penguin diagrams. In the case of
the $J/\psi K_S$ final state the weak phase of the penguin term
is the same as that of the tree contribution. Thus there is no
uncertainty for determining $\sin 2\beta$ from the CP asymmetry.

For $B^0,~\bar {B}^0 \rightarrow \pi^+ \pi^-$ the weak phases of the
tree and penguin contributions are different causing hadronic
uncertainties in the interpretation of an otherwise clean experiment.
However, by measuring also the rates of $B^0 \rightarrow \pi^0 \pi^0,~
B^+ \rightarrow \pi^+ \pi^0$ and their charge conjugate decays one can
isolate the amplitudes contributing to final states with isospin 0 and
2 and thereby determine $\alpha $ \cite{GL}. This construction, however,
relies on the fact that electoweak penguin contributions do not exist,
since they contribute to both isospins and not only to $I=0$ as the
strong penguin terms \cite{DH}. Although these weak penguin terms are
expected to be small compared to the tree amplitudes \cite{GR}, so
that the Gronau-London construction should be possible, there is still
the problem that the partial rates of the decays $B^0,~\bar {B}^0
\rightarrow \pi^0 \pi^0$ are at least an order of magnitude smaller
than for the other $2\pi$ final states \cite{KP}. In addition, because
of two neutral pions in the final state, these decays are very difficult
to measure accurately. So if this program can not be carried out the
error of $\sin 2\alpha$ is of the order of $|P/T|$, where $P~(T)$
represents the penguin (tree) contribution to $B^0 \rightarrow
\pi^+ \pi^-$.  In this connection DeJongh and Sphicas studied the behavior of
the asymmetry based on a general parameterization of the penguin magnitude and
phase \cite{DS}.

 Recently, two of us \cite{KP} calculated the effect of strong and
electroweak penguins in all $B^{\pm,0} \rightarrow \pi\pi,~\pi K$ and
$KK$ decays using specific dynamical models for the hadronic matrix
elements. Concerning the asymmetry between $\Gamma (B^0 \rightarrow
\pi^+ \pi^-)$ and $\Gamma(\bar {B}^0 \rightarrow \pi^+ \pi^-)$
 ($A_{CP}$) it turned out that the effect of electroweak penguins was
indeed small, of the order of $2\%$, and that strong penguin
amplitudes changed the asymmetry by less than $20\%$ as compared to
the tree value. These results were fairly independent of the specific
models employed for calculating the hadronic matrix elements.
Since the parameters in the time dependent asymmetry are obtained
from {\it ratios} of the weak transition matrix elements, it is clear that
they are much less model dependent than, for example, the branching
ratio. Of course, this rather moderate change of $A_{CP}$ for
$B^0 \rightarrow \pi^+ \pi^-$ depends on $P/T$ which was determined by
the model calculations. Due to the way the results in \cite{KP} were
presented only one particular set of CKM parameter values, namely
$\rho=-0.12,~ \eta=0.34$ was assumed. Although this is the preferred
value obtained in the analysis of \cite{AL} in their so called
``combined fit" it is certainly not the only possible set following
from their analysis. From CP violation in the $K^0~-\bar {K}^0$
system it is known that $\eta \neq 0$. Nevertheless, for both $\rho$
and $\eta$, only very loose bounds exist which translate into similar
loose bounds on the triangle phases $\alpha,~\beta$ and $\gamma $
\cite{AL}
\begin{equation}
  10^{\circ}< \alpha <150^{\circ}, \qquad 5^{\circ} < \beta <45^{\circ},
\qquad  20^{\circ} < \gamma < 165^{\circ}  .
\end{equation}
{}From some predictive SUSY GUT models on fermion masses and mixings,
$\alpha$ was found to be large \cite{GUT, CW}. For example, the model proposed
in \cite{CW} predicted: $\alpha = 86^{\circ}$, $\beta = 22^{\circ}$,
$\gamma = 72^{\circ}$.

It is clear that the change
in $A_{CP}$ for $B^0/\bar B^0 \rightarrow \pi^+
\pi^-$ due to penguin contributions depends not only on $P/T$ but depends also
on the particular set chosen for $\rho$ and $\eta$. However, assuming
fixed $\rho$ and $\eta$ values in \cite{KP} was an unnecessary
limitation. Of course, it would be easy to repeat the calculation of
\cite{KP} for any other set of $\rho,~\eta$ inside the bounds of (1).
This would give us a large array of numbers for $A_{CP}$. Instead we
follow in this note a different route, which is particularly simple
when we neglect the electroweak penguin terms and use some approximation for
the strong penguins.  We express the
main contribution to the asymmetry parameter, $a_{\epsilon +\epsilon'}$,
which is the coefficient of the $\sin (\Delta mt)$ term in $A_{CP}$ (see
below) in terms of the tree and penguin amplitudes and their relative
phase. Then $a_{\epsilon +\epsilon'}$ depends only on $\alpha$. This gives
us a clear insight into the dependence on $|P/T|$ and on the strong
phase and allows us to derive upper limits on the change of
$a_{\epsilon + \epsilon'}$ by including information from other decay channels
which depend on the penguin contributions more strongly than the
decay into $\pi^+ \pi^-$.

The outline of the other sections is as follows. In section 2 we give
the formulas of the asymmetry, from which we start and derive the
formula for the change due to the penguin terms. In section 3 we
present our results and discuss their relevance.

\section{CP-violating Observables in $B^{0} \rightarrow \pi^{+} \pi^{-}$}

   In this section we establish a relation between the CP-violating observables
in $B^0/\bar B^0 \rightarrow \pi^+\pi^-$, the angle $\alpha$ of the CKM
 matrix, and
an auxiliary variable $\alpha_0$ involving the ratio of penguin to tree
amplitude and the strong interaction phase difference between the tree and
the penguin amplitude.
Applying the general analysis on rephase-invariant CP-violating observables
given in ref. \cite{PW} for the B-system,
and expressing the two physical mass eigenstates $B_{L}$ and $B_{H}$ as
\begin{equation}
B_L  =  p| B^0 > + q | \bar{B^0} >, \qquad
B_H  =  p| B^0 > - q | \bar{B^0} >
\end{equation}
with the decay amplitudes of $B^0\rightarrow \pi^{+}\pi^{-}$ and
$\bar{B^0}\rightarrow \pi^{+}\pi^{-}$ written as
\begin{eqnarray}
& & g \equiv <\pi^{+}\pi^{-}|H_{eff}| B^0> = A_{T} e^{i\phi_{T} + i\delta_{T}}
+ A_{P} e^{i\phi_{P} + i\delta_{P}} \equiv \bar{h}, \\
& & h  \equiv <\pi^{+}\pi^{-}|H_{eff}| \bar{B}^0> = A_{T} e^{-i\phi_{T}
+ i\delta_{T}} + A_{P} e^{-i\phi_{P} + i\delta_{P}} \equiv \bar{g}
\end{eqnarray}
the time-evolution of states with initially pure $B^{0}$ and $\bar{B}^{0}$
 are found to be

\begin{eqnarray}
 &\Gamma(B^{0} (t) \rightarrow \pi^{+}\pi^{-} ) \propto
 \frac{1}{1 + a_{\epsilon}} \frac{(|g|^2 + |h|^2)}{2} e^{-\Gamma t}
 [ (1+ a_{\epsilon} a_{\epsilon'} ) \cosh (\Delta\Gamma t)&  \nonumber \\
&~~~~~~~+ (1+a_{\epsilon \epsilon'}) \sinh (\Delta\Gamma t) + (a_{\epsilon}
 + a_{
\epsilon'} ) \cos (\Delta m t) + a_{\epsilon + \epsilon'} \sin (\Delta m t) ]&
 \\
&&\nonumber \\
&\Gamma(\bar{B}^{0} (t) \rightarrow \pi^{+} \pi^{-} )
\propto \frac{1}{1- a_{\epsilon}} \frac{(|g|^2 + |h|^2)}{2}
 e^{-\Gamma t} [ (1+ a_{\epsilon} a_{\epsilon'} ) \cosh (\Delta\Gamma t)&
\nonumber  \\
&~~~~~~~+(1+ a_{\epsilon \epsilon'}) \sinh (\Delta\Gamma t) - (a_{\epsilon} +
a_{\epsilon'} ) \cos (\Delta m t)  -  a_{\epsilon + \epsilon'}
\sin (\Delta m t) \  ]&
\end{eqnarray}

where $a_{\epsilon}$, $a_{\epsilon'}$, $a_{\epsilon + \epsilon'}$
and $a_{\epsilon \epsilon'}$ are rephase-invariant observables and
defined as follows
\begin{eqnarray}
a_{\epsilon} & = &  \frac{1 - |q/p|^2}{1 + |q/p|^2} =
\frac{2 Re \epsilon_{B}}{1 + |\epsilon_{B}|^2} \ ,  \qquad
a_{\epsilon'} =  \frac{1 - |h/g|^2}{1 + |h/g|^2} =
\frac{2 Re \epsilon_{B}'}{1 + |\epsilon_{B}'|^2} \ ; \nonumber \\
a_{\epsilon + \epsilon'} & = & \frac{-4 Im(qh/pg)}{(1+|q/p|^2)(1+|h/g|^2)}
= \frac{2 Im\epsilon_{B} (1-|\epsilon_{B}'|^2) + 2 Im\epsilon_{B}'
(1-|\epsilon_{B}|^2) }{ (1 + |\epsilon_{B}|^2 )(1+|\epsilon_{B}'|^2)}  \\
a_{\epsilon \epsilon'} & = & \frac{4 Re(qh/pg)}{(1+|q/p|^2)(1+|h/g|^2)} -1
= \frac{4 Im\epsilon_{B} \  Im \epsilon_{B}' - 2
(|\epsilon_{B}|^2 + |\epsilon_{B}'|^2) }{( 1 + |\epsilon_{B}|^2)
( 1 + |\epsilon_{B}'|^2)} \nonumber
\end{eqnarray}
with $\epsilon_{B} = (1-q/p)/(1+q/p)$ and $\epsilon_{B}' =(1-h/g)/(1+h/g)$.
In the B system,  since  $a_{\epsilon} \ll 1$, $|\Delta \Gamma |
\ll |\Delta m |$ and $|\Delta \Gamma /\Gamma | \ll 1$ , the time-dependent
asymmetry $A_{CP}(t)$ can be simply written
\begin{equation}
A_{CP}(t) = \frac{\Gamma (B^{0} \rightarrow f) - \Gamma (\bar{B}^{0}
\rightarrow f)}{\Gamma (B^{0} \rightarrow f) + \Gamma (\bar{B}^{0}
\rightarrow f)} \simeq a_{\epsilon'}  \cos (\Delta m t) +
a_{\epsilon + \epsilon'}   \sin (\Delta m t)
\end{equation}
The CP-violating phase is related to the observables via \cite{PW}

\begin{equation}
\sin (2(\phi_{M} + \phi_{A})) =
\frac{a_{\epsilon + \epsilon'} }{\sqrt{(1-a_{\epsilon}^{2})(1 -
a_{\epsilon'}^{2})}}
\end{equation}
where the phase $\phi_{M}$ and $\phi_{A}$ are defined by
\begin{equation}
\frac{q}{p} = |\frac{q}{p}| e ^{-2i \phi_{M}}, \qquad
\frac{h}{g} = |\frac{h}{g}| e ^{-2i \phi_{A}}
\end{equation}

The tree amplitude is proportional to $v_u$ whereas the
penguin amplitude depends in general on $v_u$ and $v_c$,
where $v_{u} = V_{ub}V_{ud}^{\ast}$,
$v_{c} = V_{cb}V_{cd}^{\ast}$
 and $v_{t} = V_{tb}V_{td}^{\ast}$.
  It is well
known that when the difference of the $u$ and $c$ contributions in the
$q \bar q$ intermediate states can be
neglected, the penguin amplitude can
be expressed in terms of $v_t$ alone.  A general analysis of these
considerations has been carried through by Buras and Fleischer \cite{BF}.
In the model to be considered in the
next section this
is only violated by the additional $O(\alpha_s)$ and $O(\alpha)$ corrections
in the short distance coefficients \cite{KP}.    In this approximation,
for $\bar B^{0}\rightarrow  \pi^{+}\pi^{-}$ decay,
we have
\begin{eqnarray}
h &=& v_u \tilde T + v_t \tilde P \nonumber\\
  &=&|v_u| T e^{-i\gamma} + |v_t|P e^{i\beta}
\end{eqnarray}
Then we have
\begin{eqnarray}
  \phi_{M}&=& \beta \nonumber \\
  \phi_{T}&=& \gamma \nonumber \\
 \phi_{P} &=& -\beta
\end{eqnarray}
where $\alpha$, $\beta$ and $\gamma$ are three angles of the unitarity triangle
of the CKM matrix and $\alpha = \pi-\beta-\gamma$.
Factoring the phase $\phi_T$ of the tree contribution out we introduce the
phase shift due to the penguins, $\alpha_0$, defined by:
\begin{equation}
\phi_A=\phi_T - \alpha_0
\end{equation}
As a result we can write

\begin{equation}
\frac{a_{\epsilon + \epsilon'} }{\sqrt{(1-a_{\epsilon}^{2})(1 -
a_{\epsilon'}^{2})} } = - \sin (2(\alpha + \alpha_{0})) \simeq a_{\epsilon +
\epsilon'}
\end{equation}
where $\alpha_{0}$ is given by
\begin{eqnarray}
\tan 2\alpha_{0} & = & \frac{2 (\frac{A_{P}}{A_{T}}) \sin \Delta \phi \  \cos
\Delta \delta + (\frac{A_{P}}{A_{T}})^{2} \sin (2\Delta \phi) }{1 +
2 (\frac{A_{P}}{A_{T}}) \cos \Delta \phi \  \cos\Delta \delta
+ (\frac{A_{P}}{A_{T}})^{2} \cos (2\Delta \phi) } \nonumber \\
& = &  \frac{2 (\frac{A_{P}}{A_{T}}) \sin\alpha \  \cos \delta -
(\frac{A_{P}}{A_{T}})^{2} \sin (2\alpha) }{1 -
2 (\frac{A_{P}}{A_{T}}) \cos\alpha \  \cos\delta
+ (\frac{A_{P}}{A_{T}})^{2} \cos (2\alpha) }
\end{eqnarray}
with $\Delta \phi \equiv \phi_{T} - \phi_{P} = \pi - \alpha $
the weak phase
difference and $\Delta \delta \equiv \delta_{T} - \delta_{P} \equiv \delta $
the strong phase difference between tree and penguin diagrams.

When the strong phase $\delta$ is zero, this equation simplifies to:
\begin{equation}
\tan \alpha_0 = \frac{A_P}{A_T} \frac{\sin \alpha}{ (1-\frac{A_P}{A_T}
\cos \alpha)}
\end{equation}

(14) is a relation between the asymmetry $a_{\epsilon+\epsilon'}$, the
unitarity triangle angle $\alpha$, and the penguin complication represented by
$\alpha_0$.  (15) shows how the angle $\alpha_0$ depends on the size of the
penguin, the strong phase $\delta$ and the unitarity angle $\alpha$ itself.  To
determine $\alpha$ from the $B^0/\bar B^0 \rightarrow \pi^+\pi^-$ time
dependent
 asymmetry, $\alpha_0$
must be calculated from a model or estimated from some other process.  This
is the subject of the next section.

\section{Extraction of $\alpha$ }

  The three angles $\alpha$, $\beta$ and $\gamma$ are defined as
\begin{equation}
\alpha = arg \left( - \frac{V_{td}V_{tb}^{\ast}}{V_{ud}V_{ub}^{\ast}} \right),
\qquad \beta = arg \left( - \frac{V_{cd}V_{cb}^{\ast}}{V_{td}V_{tb}^{\ast}}
\right), \qquad
\gamma = arg \left( - \frac{V_{ud}V_{ub}^{\ast}}{V_{cd}V_{cb}^{\ast}} \right),
\end{equation}
They are related to the Wolfenstein parameters $\rho$ and $\eta$ as follows:
\begin{equation}
\tan \alpha = \frac{\eta}{\eta^{2} - \rho (1- \rho)}, \qquad
\tan \beta = \frac{\eta}{1- \rho}, \qquad \tan \gamma =
\frac{\eta}{\rho}
\end{equation}
{}From the present experimental data on $|V_{ub}/V_{cb}|$, one has
\cite{AL}
\begin{equation}
\sqrt{\rho^{2} + \eta^{2}} = 0.36 \pm 0.08 \equiv B
\end{equation}
It is clear from (14) that to extract $\alpha$ from experiment, one has to
know $\alpha_{0}$.  As shown in (15), $\alpha_{0}$ depends on $\alpha$ (or
$\eta$ and $\rho$) as well as
the ratio of the penguin amplitude $A_{P}$
to the tree amplitude $A_{T}$.
We can separate the CKM matrix elements and pure
hadronic matrix elements in the amplitudes $A_{P}$ and $A_{T}$ in the following
way,
\begin{equation}
A_{T} = |v_{u}|\ T, \qquad A_{P}=|v_{t}|\  P
\end{equation}
(20) is correct when we neglect the difference between the $u$ and $c$ quark
contributions to the penguin amplitude, discussed further below.
 In this same approximation, the
strong phase $\delta$ is zero in our model; then in this limit
(15) can be further
simplified to:

\begin{equation}
\tan \alpha_{0} = \frac{\sqrt{B^{2}-\rho^{2}}}
{B^{2} - (B^{2} - \rho)P/T}\   \left( \frac{P}{T} \right)
\end{equation}
In general, calculating the strong phase $\delta$ is difficult due to
unknown nonperturbative effects.
In \cite{KP} the strong phases
derive from absorptive parts of the $q \bar q$
intermediate states in the strong penguin contributions which are estimated
perturbatively using recently developed next to leading log formalism
 following the method pioneered by \cite{Bander}.  In \cite{KP}
it is found that
\begin{equation}
\delta_{P} \simeq 9.5^{\circ}, \qquad  \delta \simeq -9.5^{\circ},
 \qquad \delta_{T} = 0
\end{equation}
For a general consideration, we take $\delta$ as a free parameter.
The ratio $(P/T)$ is purely determined by the hadronic matrix
elements. In the operator product expansion approach, the hadronic matrix
elements are products of short-distance parts, i.e. Wilson coefficients,
evaluated by perturbative QCD, and a badly known long-distance part.
  In a factorization
approximation the long distance hadronic matrix elements are themselves
products of current form factors and decay (coupling) constants.
For the $B^{0} \rightarrow \pi^{+} \pi^{-}$ decay, it is easy to see
that the ratio $(P/T)$ is almost independent of the
uncertainties in the long-distance modeling
because differences in the approaches cancel in the ratio.
 Therefore, $\alpha_{0}$ can be well determined in a rather model
 independent way
and given by coefficients $a_i$ which have been defined in \cite{KP} in terms
of the effective Wilson coefficients $c_i^{eff}$:
\begin{equation}
a_i = c_i^{eff} +\frac{1}{N} c_j^{eff}
\end{equation}
where (i,j) is any of the pairs (1,2), (3,4), (4,5), (6,7), (7,8) and
(9,10) and N is the number of colors. (See \cite{KP} for further details.)
The second term in (23) arises from the
Fierz rearrangements in connection with the factorization
contributions.  In \cite{KP} two models were considered, $N=\infty$ and
$N=2$, to account for possible non-factorizable contributions.  Then as one
can see from Tab.\,1a,b of \cite{KP} the ratio $|P/T|$ becomes:

\begin{equation}
|\frac{P}{T}| = \frac{|a_{4} + a_{10} + (a_{6} + a_{8})R[\pi^{-},
\pi^{+}]|}{|a_{2}|}
\end{equation}
As we see from (24) this ratio does not depend on the current form factors and
decay constants.  It depends only on the factorization hypothesis and
on the effective short distance coefficients.
Furthermore it is found that the ratio is not sensitive to the
effective color number N in (23).  The reason for the simple structure of
$|P/T|$ as given by (24) is that for $\pi^+\pi^-$ states there is only one way
to factorize the transition matrix element.

The tree and penguin amplitude for $B^0 \rightarrow \pi^+\pi^-$ were
evaluated in \cite{KP} for various hypotheses concerning the $O(\alpha_s)$
corrections in the $c_i^{eff}$ (which include the absorptive parts) and the
influence of the electroweak penguins.  From these results we can calculate
the
amplitudes T and P with the CKM phases factored out.
Since the current matrix elements cancel in the ratio
this is equivalent to evaluating
P/T in terms of the coefficients $a_i$.  The results are displayed
in Tab.\,1 for N=2 and N=$\infty$.
The notation $P^x_y$ refers to penguin amplitudes arising
from $x= st$ (strong) or $x=ew$ (electroweak) penguins, and $y=u$ or $y=c$
parts of the weak Hamiltonian with and without $O(\alpha_s)$ terms in
$c_i^{eff}$, where the absorptive parts are contained in the $O(\alpha_s)$
corrections.
In the following we shall use these results in order to calculate
$\alpha_0$ for various assumptions concerning $O(\alpha_s)$ terms in the
$c_i^{eff}$ or electroweak penguin effects.  The simplest case is to use
the tree and strong penguin amplitudes with $O(\alpha_s)$ corrections
neglected, i.e. the penguin amplitudes of column 3 and 4 in Tab.\,1
which results in $\delta=0$.  The relation between these amplitudes and the
penguin
introduced above is $P= -\frac{1}{2}(P^{st''}_u +P^{st''}_c)$.
For this case $\frac{P}{T}$ in (24) is 0.05 for both
N=2 and N= $\infty$.  For fixed $B$ (21) determines $\alpha_0$ as a function
of $\alpha$ using (18).

In Fig.\,1 $\alpha_0$ is plotted as a function of $\alpha$ with $B=(\rho^2
+\eta^2)^{\frac{1}{2}} =0.28, 0.36, 0.44$, respectively, for $\delta=0$.  From
Fig.\,1 it is apparent that the penguin shift decreases with increasing $B.$
  The maximum of
$\alpha_0$ as a function of $\alpha$ occurs near $\rho=0$.  In the plot $\rho$
varies with $\alpha$ starting from $\rho<0$ to $\rho>0$ with increasing
$\alpha$.  Fig.\,1 contains our main result.  As one can see the penguin shift,
$\alpha_0$, is small compared to $\alpha$.  $\frac{\alpha_0}{\alpha}$ is
largest at small $\alpha$ and decreases monotonically with increasing $\alpha$
to zero.

To see the influence of a strong phase $\delta$ we have repeated the
calculation of $\alpha$ now using (15) where we still use the same ratio of
P/T from column 3 and 4 in Tab.\,1.
In Fig.\,2. the angle $\alpha_0$ is plotted as
a function of $\alpha$ for B=0.36 and four values of the strong phase,
$\delta=0 ^\circ, 10 ^\circ, 40 ^\circ, 90 ^\circ$. The maximal distortion
occurs for $\delta=0^\circ$ as is evident already from (15).  The behavior for
$90^\circ <\delta<180^\circ$ is approximately a reflection $\alpha_0(\delta) =
-\alpha_0(\pi-\delta)$.  $\alpha_0(\delta)$ is an even function of $\delta$.

The relation between $\alpha$ and the measured angle $\alpha_{M}$ is
$\alpha \equiv \alpha_{M}- \alpha_0$, which can be read off from Fig.\,3.
This plot is for $\delta=10 ^\circ$ and B= 0.28, 0.36, 0.44.  As we can see,
 $\alpha$ as a function of $\alpha_M=\alpha + \alpha_0$ which comes from
 the measurement
of $a_{\epsilon+\epsilon'}$ is rather independent of B and differs only
slightly from the straight line of slope unity it would be, if there were no
penguin contributions.

The results for $\alpha_0$ presented so far are for
Wilson coefficients in which $O(\alpha_s)$ and electroweak contributions
have been neglected.  The $O(\alpha_s)$ terms yield deviations from (20)
and additional absorptive contributions which generate the phase in (22).
In Tab.\,1 the penguin terms with the $O(\alpha_s)$ terms included
are given in the fifth and sixth columns as $P_u^{st}$ and $P_c^{st}$.
As we can see the $O(\alpha_s)$ terms change the penguin terms up to 24\%
and produce the deviation $$\frac{P_u^{st}-P_c^{st}}{P_u^{st}+P_c^{st}}=0.13$$
in the real parts.  In addition there is an imaginary part of the same order
as the real part (these numbers are for the N=2 case).  Of course the resulting
shift in $\alpha_0$ depends on $\alpha$ or equivalently on the value of $\rho$.
Instead of calculating $\alpha_0$ as a function of $\alpha$ (or $\rho$) we
quote only results where $\alpha$ nearly had its maximum, i.e., $\rho = 0$,
at $\alpha= 70^\circ$.
By calculating $a_\epsilon'$ and $a_{\epsilon + \epsilon'}$ directly we can use
(14) to extract $\alpha_0$ for this value of $\alpha$.  The major effect is
due to the strong penguin amplitude itself, without $\alpha_s$ or
electroweak
corrections, which shifts $\alpha$ by $\alpha_0=8.0^\circ$.  The $\alpha_s$
and electroweak corrections shift $\alpha$ by an additional $\Delta \alpha =
1.9^\circ$ and $0.6^\circ$ respectively, to a total shift of
$\alpha_0=10.5^\circ$.

The values of the strong phase may be extracted from the direct
CP-violating parameter $a_{\epsilon'}$. Hence in principle one can determine
$\delta$ without recourse to a model.  On general grounds, however, one would
expect $\delta$ to be small.

It is clear that the shift $\alpha_0$ due to the penguin effects is small.
For all $\alpha$ the relative shift $\alpha_0/\alpha$ is less than $30\%$
and decreases from this value
with increasing $\alpha$.  This result depends on our model of the penguin
amplitudes, in particular on $P/T$.  In principle one need not rely on the
model but rather obtain $|P|$ from pure penguin or penguin dominated
processes, i.e., $\bar K^0 \pi^-$ (pure penguin) and $K^+\pi^-$
(penguin dominated).  Unfortunately there are only experimental upper bounds
on the branching ratios of these decays.  In reference
\cite{KP} we have calculated the branching
ratios of these decays for the special choice $\rho=-0.12, \eta=0.34$.  Since
$\bar K^0 \pi^-$ depends only on $|V_{ts}|$ which is well known, we can obtain
 upper limits on $|P|$. Taking for example the N=2 case, when we compare with
the experimental limit \cite{Asner},
$BR (B^- \rightarrow \bar K^0 \pi^-)<4.8 \times 10^{-5}$, we find that
 $|P_{exp}/P|<2.2.$  Keeping T fixed, such a large value of the penguin
 amplitude
given by the upper limit of $|P_{exp}|$
would increase $\alpha_0$ from $8^\circ$ to $17^\circ$ for the maximal shift
 at $\rho=0$.  Here we make the assumption that the penguin amplitudes
of $K\pi$ and $\pi\pi$ final states are related, i.e. a larger $P$
in $K\pi$ would mean a larger $P$ in $\pi\pi$.  This can be justified with
SU(3) symmetry arguments.  SU(3) symmetry breaking effects are indeed moderate
in our model calculations.

The penguin dominated decay channel $\bar B^0 \rightarrow
 K^+\pi^-$ gives us a much better limit as advocated by Silva and
Wolfenstein \cite{Wolf}.  The experimental limit
 on this branching
ratio is $1.7 \times 10^{-5}$;  from our previous work we know that this
decay is dominated by the penguin amplitude (since the tree amplitude is
Cabibbo suppressed) in the ratio 4:1 in the amplitude.  This gives
about $|P_{exp}/P|<1.4$ leading to an even smaller shift of $\alpha_0$ to
$\simeq 11^\circ$ compared to the $8^\circ$ calculated above.  Other
 experimental
limits on the branching ratios relevant for comparing with our model are
$BR(\pi^-\pi^0)<1.6 \times10^{-5}$,
$BR(\pi^+\pi^-)<2.0 \times 10^{-5}$ and $BR(\pi^+\pi^- + K^+\pi^-)=
(1.8\pm0.6)\times 10^{-5}$
\cite{Asner}.  Our model obeys these constraints; in particular we obtained
$BR(\pi^+\pi^- + K^+\pi^-) = 2.15 \times 10^{-5}$ for $\rho=0, \eta=B$ which
agrees perfectly with the measured value.  This shows that the tree and
penguins can not be too far from the values in our model.

 In conclusion, we find that the penguin distortion in the determination of
$\alpha$ from the $\pi^+\pi^-$ asymmetry is not a real obstacle provided
$\alpha$ is not too small.  Even if the penguin amplitude is taken from the
upper limit of the pure penguin dominated process the penguin distortion on
$\alpha$ remains below $\sim25\%$ at large $\alpha$.  Improved experimental
work on exclusive charmless hadronic B decays will even more sharply constrain
the size of penguin amplitudes and in turn limit the shift $\alpha_0.$


\def\i{i}                  
\def\as{\mbox{$\alpha_s$}} 
\def\Oa{$O(\alpha_s)$}     
\def\Oaa{$O(\alpha_s^2)$}  
\def\msb{$\overline{ms}$~} 
\def\to{\rightarrow}       
\def\Im{Im}
\def\Re{Re}
\def\ln{\mbox{$\ell n$}}   
\def\sin{\mbox{ sin}}
\def\cos{\mbox{ cos}}
\def\Sum{\displaystyle\sum}
\def\eff{{\rm eff}}
\def\Heff{{\cal H}_{\rm eff}}
\def\ce{c^{\eff}}
\def\me{{\bf m}}
\def\r{{\bf r}}

\def\JPsi{J/\Psi}

\def\eq{}

\def\bra{\langle}
\def\ket{\rangle}
\eject
\noindent {\bf Table Caption}
\\
\\
{\bf Tab.\,1}: Reduced amplitudes for $\bar B^0 \rightarrow \pi^+\pi^-$.
CKM phases have
been removed.  These amplitudes are based on a next to leading log corrected
weak Hamiltonian with $\alpha_s$ and $\alpha$ corrections, assuming
factorization of the hadronic matrix elements.  Numbers in the parenthesis
are the real and imaginary part of the amplitude. For further details, see
reference \cite{KP}.

\eject
\centerline{Tab.\,1}

{\small

\begin{center}
\begin{tabular}{||l||l||l|l|l|l|l|l|l||}
\hline\hline
\multicolumn{8}{||c||}{{Reduced Amplitudes $\bar B^0 \rightarrow
\pi^+\pi^-$}}\\
\multicolumn{8}{||c||}{Tree, Strong and EW Penguins $T$, $P^{st}$, $P^{ew}$
 with () or
without ('')
    $\alpha_s$ corrected} \\
\multicolumn{8}{||c||}{{ NLL QCD Coefficients}}\\
\hline
\hline
N     & T &${P^{st}_u}''$&${P^{st}_c}''$&$P^{st}_u$ &$P^{st}_c$
                                            &$P^{ew}_u$&$P^{ew}_c$\\
\hline \hline
$\infty$ & 2.84&(-0.141,0) &(-0.141,0)&(-0.134,-0.0548)&(-0.175,-0.0291)
                                         &(0.0052,0) &(0.0052.0) \\
$2$      & 2.44&(-0.115,0) &(-0.115,0) &(-0.109,-0.0457)&(-0.143,-0.0243)
                                         &(0.0052,0) &(0.0052,0) \\
\hline \hline
\end{tabular}
\end{center}   }
\eject
\newpage

\noindent{\bf Figure Caption}
\\
\\
{\bf Fig.\,1.}  $\alpha_{0}$ as a function of $\alpha$ with
$\delta = 0^{\circ}$.
The curves correspond to
$B = 0.28$ (dotted), $0.36$ (solid), and $0.44$ (dashed).
All angles are in degrees.
\\
\\
{\bf Fig.\,2.}  $\alpha_{0}$ as a function of $\alpha$ with
$B=0.36$. The curves correspond to the strong phase
$\delta =0^{\circ}$ (dashed),  $10^{\circ}$
(solid), $40^{\circ}$ (dotted), and $90^{\circ}$ (dot-dashed).  All angles
are in degrees.
\\
\\
{\bf Fig.\,3.}  $\alpha = \alpha_{M} - \alpha_{0}$ as a function of
the measured $\alpha_{M}$ with $\delta = 10^{\circ}$.
The curves correspond to
$B = 0.28$ (dotted),  $0.36$ (solid), and $0.44$ (dashed).  All angles
are in degrees.

\end{document}